# Massive Connection Bosons

Gustavo R. González-Martín[a]

Dep. de Física, Universidad Simón Bolívar, Apdo. 89000, Caracas 1080-A, Venezuela.

It is shown that geometric connection field excitations acquire mass terms from a geometric background substratum related to the structure of space-time. Commutation relations in the electromagnetic $su(2)_Q$ sector of the connection limit the number of possible masses. Calculated results, within corrections of order $\alpha$, are very close to the masses of the intermediate W, Z bosons.

04.50+h, 14.70Fm, 14.70Hp, 02.20.Qs

[a] Webpage http://prof.usb.ve/ggonzalm

# Introduction

In the standard model of particle theory it is accepted that interaction fields may acquire mass by the Higg's mechanism.[1] On the other hand, many properties of particle are related to the space-time structure of special relativity. Therefore it should not be a great surprise that there is a curved geometry [2], fundamentally associated to the space-time structure of general relativity, whose field equations determine a particular substratum solution that provides a constant mass parameter for the linear Yukawa [3] equations obeyed by its excitations. The connection excitations may represent geometric bosonic particles.

The geometry is related to a connection $\Gamma$ in a principal fiber bundle $(E,M,G)$. The structure group $G$ is $SL(4,\mathbb{R})$ and the even subgroup $G_+$ is $SL_1(2,\mathbb{C})$. The subgroup $L$ (Lorentz) is the subgroup of $G_+$ with real determinant, in other words, $SL(2,\mathbb{C})$. Electromagnetism is associated to an $SU(2)_Q$ subgroup. Matter is represented by a current $J$ expresed in terms of a principal fiber bundle section $e$, which is related by charts (coordinates) to matrices of the group $SL(4,\mathbb{R})$. This structure defines a frame of $SL(4,\mathbb{R})$ spinors over space-time. The connection is an $sl(4,\mathbb{R})$ 1-form that acts naturally on the frame $e$ (sections).

It has been shown [4] that fermionic particle mass ratios may be obtained using a substratum solution and related symmetric subspaces of the coset $K = G/G_+$ . We have chosen that the substratum matter be referred to itself. In this manner, the local matter frame $e_b$, referred to the reference frame $e_r$ becomes the group identity $I$. Actually this generalizes comoving coordinates (coordinates adapted to dust matter geodesics).[5] We adopt coordinates adapted to local substratum matter frames (the only non arbitrary frame is matter itself, as are the comoving coordinates). If the frame $e$ becomes the identity $I$, the substratum current density becomes a constant.

At the small distance $\lambda$, characteristic of excitations, the elements of the substratum, both connection and frame, appear symmetric, independent of space time. These are the necessary conditions for the substratum to locally admit a maximal set of Killing vectors [6] of the space-time symmetry of the connection (and curvature). This means that there are space-time Killing coordinates such that the connection is constant non zero in the small region of particle interest. A flat connection does not satisfy the field equation. The excitations may always be taken around a symmetric non zero connection. If we look at the left side of the field equation [2] we notice that for a constant connection form $\omega$, the expression reduces to triple wedge products of $\omega$ with itself, which may be put in the form of a polynomial in the components of $\omega$. This cubic polynomial represents a self interaction of the connection field since it may also be considered as a source for the differential operator. Substratum solutions of this field equation are discussed in the appendix. They provide constant parameters for the fluctuation equations of the geometric field equation.

# Particular Excitation Solution

In particular, we assume a connection excitation with algebraic components only in the complex Minkowski plane $K_k$. These excitations exist around a constant odd complex solution which was constructed in the appendix by extending the real substratum solution to complex functions. The connection form, in terms of the geometrical fundamental unit of length is

$$\omega = \omega^\pm + \delta\omega = -m_g dx^{\hat{\alpha}\pm}\left(\kappa_{\hat{\alpha}} \pm i\kappa_{\hat{\alpha}}\right) + \delta\omega , \tag{1}$$

$${}^*d\,{}^*d\delta\omega_\delta^\alpha + \left|\omega^\pm\right|^2 \delta\omega_\delta^\alpha + 2\left|\omega^\pm\right|^2 \delta\omega_\rho^{\hat{\rho}}\delta_\delta^\alpha = 4\pi\alpha\,\delta J_\delta^\alpha \delta\left(\mathbf{x}-\mathbf{x}'\right) . \tag{2}$$

Given a representation, a solution to these coupled linear equations may always be found in terms of the Green's function of the differential operator and the current excitation $\delta J$, which may be an extended source. In order to decouple the equations, it is necessary to assume that $\delta\omega_\rho^{\,\rho}$ vanishes. If this is the case then all equations are essentially the same and the solution is simplified.

We shall restrict ourselves to consider the equation for a unit point excitation of the ***odd current*** and its solution which is the Green's function. If we assume a time independent excitation with spherical symmetry, the only relevant equation would be the radial equation. Since $\omega^\pm$ is constant, it represents an essential singularity of the differential equation, an irregular [3] singular point at infinity. The corresponding solutions have the exponential Yukawa behavior. Let us interpret that the absolute value $\omega^\pm$ of the curved substratum gives an effective range $\omega^{-1}$ to the linear excitations. Equation (2) time independent for a point source is, designating the fluctuation as a weak field $W$,

$$\nabla^2 W(\mathbf{x}) - \omega^2 W(\mathbf{x}) = -4\pi\alpha^2 \left(\frac{e}{\bar{e}}\right)^2 \delta(\mathbf{x}-\mathbf{x}') \equiv -4\pi\alpha^2 g^2 \delta(\mathbf{x}-\mathbf{x}') , \tag{3}$$

where we explicitly recognize that the current fluctuation is of order $\alpha$ or equivalently of order charge squared. Furthermore, we realize that the assumed excitation is the odd part of an $su(2)_Q$ representation and should explicitly depend on the odd part of the charge. The geometric charge that enters in this first order perturbation equation is the $su(2)_Q$ charge defined by the original nonlinear unperturbed equation, which is the electric charge quantum $e=1$. We should define a $g$ factor that expresses this geometric charge quantum in terms of its odd component, as a unit of a weak charge. The factor $(\alpha g)^2$ is not part of the $\delta$ function which represents a unit odd charge. This equation may be divided by $(\alpha g)^2$, obtaining the equation for the odd excitations produced by the unit point odd charge or equation for the Green's function,

$$\frac{1}{(\alpha g)^2}\nabla^2 W(x) - \left(\frac{\omega}{\alpha g}\right)^2 W(x) = -4\pi\delta(x-x') . \tag{4}$$

We may define a new radial coordinate $r=\alpha g x^i$, rationalized to a new range $^{-}\mu^{-1}$ for odd excitations, which incorporates the $\alpha g$ constant. Taking in consideration the covariant transformations of $W$ and $\delta$, the radial equation is then

$$\frac{1}{r}\frac{\partial^2}{\partial r^2}(rW) - {^-\mu^2}W = -4\pi\delta(r-r') . \tag{5}$$

Since $^{-}\mu$ is constant, the Green's function for this differential operator is

$$\mathcal{G} = \frac{1}{4\pi}\frac{e^{-{^-\mu}|x-x'|}}{|x-x'|} = \frac{1}{4\pi}\frac{e^{-{^-\mu}r'}}{r'} \tag{6}$$

whose total space integration introduces, in general, a range factor

$$\frac{1}{4\pi}\int_0^\infty dr'r'e^{-{^-\mu}r'}\int_0^{4\pi} d^2\Omega' = \frac{1}{{^-\mu^2}} . \tag{7}$$

The range $^{-}\mu$ may be evaluated using eq. (2),

$$\omega^2 = g_{CC}\left(\omega_\mu\omega^\mu\right) = \omega_\mu^*\omega^\mu = 8m_g^2 , \tag{8}$$

giving

$$^-\mu = \frac{\omega}{\alpha g} = \frac{2\sqrt{2}m_g}{\alpha g} . \tag{9}$$

## Massive SU(2) Excitations.

The $su(2)_Q$ connection excitations may be considered functions over the sphere SU(2)/U(1). In similarity with the quantization of the angular momentum, we concluded in a previous paper [7] that the $SU(2)_Q$ electromagnetic excitations have indefinite azimuthal directions but quantized polar directions determined by the possible translation values on the sphere. The internal direction of the potential $A$ must be along the possible directions of the electromagnetic generator $E$ in $su(2)_Q$ as indicated in the figure. The $A$ components must be proportional to the possible even and odd translations. In consequence, the total $A$ vector must lie in a cone defined by a quantized polar angle $\theta$ relative to an axis in the even direction and an arbitrary azimuthal angle.

The fundamental state that represents an electromagnetic quantum is the $SU(2)_Q$ state with charge +1, corresponding to the ½, ½ eigenvalues of the electromagnetic rotation. The corresponding $\theta$ angle is

$$\frac{|^-E|}{|^+E|} = \frac{\sqrt{\frac{1}{2}(\frac{1}{2}+1)}}{\frac{1}{2}} = \sqrt{3} = \tan\left(\pi/3\right) \equiv \tan\theta_{1/2}^{1/2} . \tag{10}$$

The $su(2)$ connection excitations over the bidimensional sphere SU(2)/U(1) are generated by the $^-E$ odd generators. The complex charged raise and lower generators $^-E^\pm$ are defined in terms de the real generators,

$$^-E^\pm = E^1 \pm iE^2 ,\qquad (11)$$

which raise or lower the charge $q$ of an eigen state as follows

$$^-E^\pm |c,q\rangle = N|c,(q\pm 1)\rangle . \qquad (12)$$

Therefore, these excitations, defined as representations of the $SU(2)_Q$ group, require a substratum with a preferred direction along the even quaternion $q^3$ in the $su(2)_Q$ sector. Only the complex solution discussed in the appendix, eq. (46), gives an adequate substratum. We shall call this substratum the odd complex substratum.

The odd complex substratum solutions in the $su(2)$ sector reduce to

$$^-\omega = {}^-\omega^{0\pm}\left(\kappa_0 \pm i\kappa_1\kappa_2\kappa_3\right) = {}^-\omega^{0\pm}\, {}^-q^\pm \qquad (13)$$

and should correspond to the two $^-E$ odd sphere generators. This odd $su(2)$ subspace generated by the complex external solution is physically interpreted as an $SU(2)$ electromagnetic substratum, a vacuum, with potential $^-\omega$ which determines ranges which may be interpreted as masses for the excitations of the electromagnetic connection in its surroundings. The range $^-\mu^{-1}$ is determined by $^-\omega$, the odd fundamental bidimensional component in the substratum equatorial plane.

The even and odd components of the $su(2)_Q$ excitation in the $E$ direction have to obey the relations,

$$^+A^2 + {}^-A^2 = A^2 . \qquad (14)$$

Due to the quantization of the $su(2)$ connection, there only are two possible definite absolute values, associated to a fundamental representation, which are related by

$$\frac{|^-A|}{|A|} = \frac{|^-E|}{|E|} = \sin \theta_{1/2}^{1/2} . \qquad (15)$$

An odd $su(2)_Q$ connection excitation should be the spin 1 boson ($SU(2)_S$ representation) which also is a fundamental $SU(2)_Q$ representation where the three component generators keep the quantized relations corresponding to electromagnetic rotation eigenvalues ½, ½, similar to the frame excitation representation (proton). Of course, they differ regarding the spin $SU(2)_S$ because the former is a spin 1 representation and the latter is a spin ½ representation. We shall call this excitation by the name fundamental complete excitation. A fluctuation of the odd substratum

$$\delta\left(^-\omega\right) = \delta\left(^-\omega^{0\pm}\right) {}^-q^\pm \qquad (16)$$

does not provide a complete $su(2)_Q$ excitation.

The excitation ranges are proportional to the absolute values of a connection. The only possible ranges associated with an $su(2)_Q$ excitation should be proportional to the only possible values of the quantized connection

$$\frac{|^-A|}{|A|} = \frac{^-\mu}{\mu} . \qquad (17)$$

The $su(2)_Q$ currents are similarly quantized and are related by

$$\frac{|^-j|}{|j|} = \frac{|^-E|}{|E|} = \sin \theta_{1/2}^{1/2} = \frac{|^-e|}{|e|} . \qquad (18)$$

The $g$ factor that expresses the geometric charge quantum in terms of its odd unit component is

$$g = \csc \theta_{1/2}^{1/2} \qquad (19)$$

and we obtain, omitting the polar angle indices,

$$^-\mu = \frac{\omega}{\alpha g} = \frac{2\sqrt{2} m_g \sin\theta}{\alpha} \qquad (20)$$

and

$$\mu = \frac{2\sqrt{2}m_g}{\alpha} \ . \tag{21}$$

On the other hand, the excitation ranges should be provided by the excitation field equation (2) as the absolute value of some substratum connection. The odd complex substratum admits a related substratum with an additional even component $^+\omega$ connection that should provide the value of the smaller range $\mu^{-1}$. This even part $^+\omega$ increases the modulus of the total connection. The possible value of $^+\omega$ should correspond to the allowed value of the total substratum connection absolute value $\mu$. We shall call this substratum the secondary complex substratum.

The value $\mu^2$ represents the bound energy of a complete (with its three components) $su(2)_Q$ fundamental excitation around the secondary complex substratum with equation

$$\nabla^2 A(x) - \mu^2 A = -4\pi\delta(x - x') \ . \tag{22}$$

Since the orientations of the potential $A$ and the background connection $\omega$ are quantized, their decompositions into their even and odd parts are fixed by the representation of the connection. Therefore this $\mu^2$ energy term may be split using the characteristic polar angle $\theta$,

$$\mu^2 = \mu^2\left(\cos^2\theta + \sin^2\theta\right) = (\mu\cos\theta)^2 + (\mu\sin\theta)^2 \equiv {}^+\mu^2 + {}^-\mu^2 \ . \tag{23}$$

The $^+\mu$ corresponds to the even component $^+\omega$ associated with the U(1) group generated by the $\kappa_5$ even electromagnetic generator.

This energy $\mu$ may not be decomposed without dissociating the complete su(2) excitation due to the orientation quantization of its components. If the complete excitation is disintegrated into its partial components, the corresponding equation for the even component separates from the odd sector in $K$, as indicated in eq. (2), and has an abelian connection. Therefore, no mass term appears in the even component equation, physically consistent with the zero mass of the photon. The energy associated to $^+\omega$ in the $\kappa_5$ direction is available as free energy. For a short duration the splitting takes energy from the substratum. On the other hand, $^-\mu$ corresponds to the odd complex substratum and when the complete excitation is disintegrated, the energy appears as the mass term in the coupled equations (2) associated to the pair of odd generators $^-E^\pm_\pm$.

$$\nabla^2 {}^-A(x) - \left(\mu^2 - {}^+\mu^2\right){}^-A + L_I(\delta\omega) = -4\pi\delta(x-x') \ . \tag{24}$$

We may neglect the coupling term $L_I$, as was done previously in eq. (2)), so that the equation may approximately be written

$$\nabla^2 {}^-A(x) - {}^-\mu^2 {}^-A = -4\pi\delta(x-x') \ . \tag{25}$$

The fundamental real fermionic solution that represents the proton has a mass (energy), given by eq (34) as indicated in the appendix,

$$m_p = 4m_g \ . \tag{26}$$

The odd complex substratum solution has a similar relation in terms of $m_g$, eq. (48),

$$\omega = \left|\frac{1}{4}\operatorname{tr}\omega^*_\mu\omega^\mu\right| = 2\sqrt{2}m_g = \alpha g\mu \ . \tag{27}$$

Considering these relations, the $\theta$ polar angle also determines the ratio of the two energies or masses associated to the fundamental excitation,

$$\frac{m_{{}^-A}}{m_A} = \sin\theta \ . \tag{28}$$

These relations determine that the SU(2) connection excitation energies or masses are proportional to the proton mass $m_p$,

$$m_A = \mu = \frac{2\sqrt{2}m_g}{\alpha} = \frac{m_p}{\sqrt{2}\alpha} = 90.9177 \text{ Gev} \approx m_Z = 91.188 \text{ Gev} \ , \tag{29}$$

$$m_{_A} = m_A \sin\theta = \frac{m_p \sin\theta}{\sqrt{2}\alpha} = 78.7370 \text{ Gev} \approx m_W = 80.42 \text{ Gev} . \tag{30}$$

These values indicate a relation with the weak intermediate bosons. They also indicate a possible relation of this polar angle with Weinberg's angle [8]: Weinberg's angle would be the complement of the polar angle. They admit corrections of order $\alpha$. For example if we use the corrected value for $\theta$, obtained from its relation to the experimental value of Weinberg's angle,

$$m_{_A} = \frac{m_p \sin\theta}{\sqrt{2}\alpha} = 79.719 \text{ Gev} \approx m_W = 80.42 \text{ Gev} . \tag{31}$$

Let us define a monoexcitation as an excitation that is not a complete $SU(2)_Q$ representation and is associated to a single generator. A collision may excite a resonance at the $A$ fundamental connection excitation energy $m_A$ determined by the secondary complex substratum. The fundamental connection excitation $A$ which is an $SU(2)$ representation may also be dissociated or disintegrated in its three components ($^+A$, $^-A^\pm$), when the energy is sufficient, as two $^-A^\pm$ free monoexcitations, each one with mass (energy) determined by the odd complex substratum, and a $^+A$ third free $U(1)$ monoexcitation, with null mass determined by the even complex substratum component $^+\omega$ which is abelian.

This simply determines the existence of four spín-1 excitations or boson particles associated to the $SU(2)$ connection and their theoretical mass values: 1. The resonance at energy $\mu$ of the fundamental potential $A$, which has to decay neutrally in its components ($^+A$, $^-A^\pm$) and may identified with the $Z$ particle; 2. The pair of free monoexcitations $^-A^\pm$, charged $\pm 1$ with equal masses, which may identified with the $W^\pm$ particles; 3. The free excitation $^+A$, neutral with null mass which may identified with the photon.

In this manner we may give a geometric interpretation to the $W, Z$ intermediate bosons and to Weinberg's angle which is then equivalent to the complement of $\theta_{1/2}$ polar angle.

# Appendix

## The complex substratum

It has been shown[9] that there is a real background (substratum) connection solution,

$$\omega = e^{-1}\left(-m_g dx^{\hat{\alpha}} \kappa_{\hat{\alpha}}\right) e + e^{-1} de = -m_g J + e^{-1} de , \tag{32}$$

which allows the calculation of mass quotients. The solution corresponds to the local moving frame rest system, where all derivatives are assumed zero in the non linear differential operator. The value of $m_g$ is determined by the only remaining connection contribution to the field equation left side, which is the connection triple product in the covariant derivative,

$$\begin{aligned}{}^*\left[\omega \wedge {}^*(\omega \wedge \omega) - {}^*(\omega \wedge \omega) \wedge \omega\right]^\alpha &= \delta^{\rho\alpha}_{\mu\nu} \omega^c_\rho \omega^{a\mu} \omega^{b\nu}\left[E_c, E_a E_b\right] \\ &= \omega^c_\rho \omega^{a\rho} \omega^{b\alpha}\left[E_c,\left[E_a, E_b\right]\right] = 4\pi\alpha J^\alpha .\end{aligned} \tag{33}$$

The real substratum is obtained by restricting the generators $E$ to the orthonormal set $\kappa_\alpha$. The definition of mass in terms of the Cartan-Killing metric is

$$m = \tfrac{1}{4}\mathrm{tr}\left(J^\mu \Gamma_\mu\right), \tag{34}$$

and gives a relation between the fundamental inverse length of the geometry $m_g$ and the mass of a fundamental particle, which we take as the proton,[9]

$$m_p = 4m_g . \tag{35}$$

The symmetric space $K$ has a complex structure [10]. The center of $G$, which is not discrete, contains a generating element $\kappa_5$ whose square is -1. We shall designate by $2J$ the restriction of $\mathrm{ad}(\kappa_5)$ to the tangent space $TK_k$. This space, that has for base the 8 matrices $\kappa_\alpha, \kappa_\beta \kappa_5$, is the proper subspace corresponding to the eigenvalue -1 of the operator $J^2$, or,

$$J^2\left(x^\lambda \kappa_\lambda + y^\lambda \kappa_\lambda \kappa_5\right) = \tfrac{1}{4}\left[\kappa_5,\left[\kappa_5, x^\lambda \kappa_\lambda + y^\lambda \kappa_\lambda \kappa_5\right]\right] = -x^\lambda \kappa_\lambda - y^\lambda \kappa_\lambda \kappa_5 . \tag{36}$$

The endomorphism $J$ defines a complex structure over $K$.

The real substratum connection may be extended to the complex field using this complex structure

$$\omega = \omega^a E_a + {}^+\omega \kappa_5 , \tag{37}$$

where we indicate the eight $K$ space generators by $E_a$. If we restrict the latin generator indices from 1 to 9 we may write the left side of the field equations with the single contribution of the connection triple product,

$${}^*\left[\omega \wedge {}^*(\omega \wedge \omega) - {}^*(\omega \wedge \omega) \wedge \omega\right]^\alpha = \delta^{\rho\alpha}_{\mu\nu} \omega^c_\rho \omega^{a\mu} \omega^{b\nu} [E_c, E_a E_b] = \omega^c_\rho \omega^{a\rho} \omega^{b\alpha} [E_c, [E_a, E_b]], \tag{38}$$

here indicated by $X$, in the following form,

$$\begin{aligned} X^\alpha &= \omega^c_\mu \omega^{b\mu} \omega^{a\alpha} [E_c, [E_b, E_a]] + \omega^c_\mu \omega^{b\mu} {}^+\omega^\alpha [E_c, [E_b, \kappa_5]] \\ &+ \omega^c_\mu {}^+\omega^\mu \omega^{a\alpha} [E_c, [\kappa_5, E_a]] + \omega^c_\mu {}^+\omega^\mu {}^+\omega^\alpha [E_c, [\kappa_5, \kappa_5]] \\ &+ {}^+\omega_\mu \omega^{b\mu} \omega^{a\alpha} [\kappa_5, [E_b, E_a]] + {}^+\omega_\mu \omega^{b\mu} {}^+\omega^\alpha [\kappa_5, [E_b, \kappa_5]] \\ &+ {}^+\omega_\mu {}^+\omega^\mu \omega^{a\alpha} [\kappa_5, [\kappa_5, E_a]] + {}^+\omega_\mu {}^+\omega^\mu {}^+\omega^\alpha [\kappa_5, [\kappa_5, \kappa_5]] , \end{aligned} \tag{39}$$

The total sl(4,$\mathbb{R}$) odd vector subspace, spanned by $\kappa_\alpha$ and $\kappa_5 \kappa_\beta$, is isomorphic to $K_k$, the tangent space of the symmetric coset $K$ at a point $k$. The complex structure allows us to define a complex metric in its tangent space using the Cartan-Killing metric. Thus, the total sl(4,$\mathbb{R}$) odd vector subspace may be considered as a complex Minkowski space with metric $\eta$,

$$g_{CC}(X,Y) \equiv g_C(X^*, Y) = -\eta(X^*, Y) . \tag{40}$$

Since the generator $\kappa_5$ commutes with the even sector and generates the complex structure, eq. (39) becomes

$$\begin{aligned} X^\alpha &= \omega^c_\mu \omega^{b\mu} \omega^{a\alpha} [E_c, [E_b, E_a]] + 2\omega^c_\mu \omega^{b\mu} {}^+\omega^\alpha \kappa_5 \{E_c, E_b\} \\ &- 2\omega^c_\mu {}^+\omega^\mu \omega^{a\alpha} \kappa_5 \{E_c, E_a\} + {}^+\omega_\mu \omega^{b\mu} {}^+\omega^\alpha E_b - {}^+\omega_\mu {}^+\omega^\mu \omega^{a\alpha} E_a . \end{aligned} \tag{41}$$

We may use the $K$ complex structure to take $\omega$ as a complex functions and the generators as $\kappa_\alpha$,

$$\begin{aligned} X^\alpha &= \omega^c_\mu \omega^{b\mu} \omega^{a\alpha} [\kappa_c, [\kappa_b, \kappa_a]] - 4\omega^c_\mu \omega^{b\mu} {}^+\omega^\alpha \kappa_5 \eta_{cb} \\ &+ 4\omega^c_\mu {}^+\omega^\mu \omega^{a\alpha} \kappa_5 \eta_{ca} + 4\, {}^+\omega_\mu \omega^{b\mu} {}^+\omega^\alpha \kappa_b - 4\, {}^+\omega_\mu {}^+\omega^\mu \omega^{a\alpha} \kappa_a . \end{aligned} \tag{42}$$

Finally, we evaluate the $K$ components in the term $X$,

$$\tfrac{-1}{4} \mathrm{tr}\, \kappa_d X^\alpha = \tfrac{-1}{4} \mathrm{tr}\, \kappa_d \left( \omega^c_\mu \omega^{b\mu} \omega^{a\alpha} [\kappa_c, [\kappa_b, \kappa_a]] + {}^+\omega_\mu \omega^{b\mu} {}^+\omega^\alpha \kappa_b - {}^+\omega_\mu {}^+\omega^\mu \omega^{a\alpha} \kappa_a \right), \tag{43}$$

$$X^\alpha_d = -4 \left( \omega^{c*}_\mu \omega^\mu_c \omega^\alpha_d - \omega^{c*}_\mu \omega^\mu_d \omega^\alpha_c \right) - 4 \left( {}^+\omega_\mu {}^+\omega^\mu \omega^\alpha_d - {}^+\omega_\mu \omega^\mu_d {}^+\omega^\alpha \right) \tag{44}$$

and, using this expression for the left side $X$ write the field equation as

$$\left( \left( \omega^{c*}_\mu \omega^\mu_c \omega^\alpha_d - \omega^{c*}_\mu \omega^\mu_d \omega^\alpha_c \right) + \left( {}^+\omega_\mu {}^+\omega^\mu \omega^\alpha_d - {}^+\omega_\mu \omega^\mu_d {}^+\omega^\alpha \right) \right) \kappa^d \\ + \left( \omega^{c*}_\mu \omega^\mu_c {}^+\omega^\alpha - \omega^{c*}_\mu {}^+\omega^\mu \omega^\alpha_c \right) \kappa_5 = -\pi \alpha \delta J^\alpha . \tag{45}$$

If we assume that the even connection component ${}^+\omega$ is zero we obtain the odd complex substratum solution which has the form,

$$\omega = -m_g dx^{\hat{\alpha}\pm} \left( \kappa_{\hat{\alpha}} \pm \kappa_5 \kappa_{\hat{\alpha}} \right) \equiv -m_g (1 \pm i) \kappa_{\hat{\alpha}} dx^{\hat{\alpha}\pm} , \tag{46}$$

with the same value for the constant $m_g$ given by eq. (32).

We interpret this complex connection solution as a complex substratum with a complex matter current.

$$J^\lambda = \left((1+i)\kappa^\lambda - (1+i)\overline{\kappa}^\lambda\right) = 2(1+i)\kappa^\lambda \ . \tag{47}$$

We may represent the odd complex substratum connections $\omega^\pm$ given by this solution as points on the complex Minkowski space $K_k$, with equal moduli,

$$\left|\omega^\pm\right| = \sqrt{\eta\left(\omega^{\pm *}\cdot\omega^\pm\right)} = \sqrt{-g_C\left(\omega^{\pm *}\cdot\omega^\pm\right)} = \sqrt{-g_C\left(\omega^+\cdot\omega^-\right)}$$
$$= \sqrt{\tfrac{-1}{4}\,\mathrm{tr}\left(m_g^2\left(\kappa_{\hat\alpha}\kappa^{\hat\alpha} + \kappa_5\kappa_{\hat\alpha}\kappa_5\kappa^{\hat\alpha}\right)\right)} = 2\sqrt{2}\,m_g \ . \tag{48}$$

The points define complex vectors $\bar{q}^\pm$,

$$\bar{q}_\alpha^+ = \kappa_\alpha + i\kappa_5\kappa_\alpha \equiv \frac{\partial}{\partial x^\alpha} + i\frac{\partial}{\partial y^\alpha} = \frac{\partial}{\partial z^\alpha} \tag{49}$$

and their complex conjugates, which form a base on the $K_k$ sector, related to the associated complex coordinates $z^\alpha$. This pair of vectors together with the vector determined by the static connection $^+\omega$,

$$^+q^3 = \kappa_5 \ , \tag{50}$$

provide a quaternion base on a compact subspace for the subalgebra $su(2)_Q$.

If we assume that the connection even component $^+\omega$ is non zero we should extend the source current accordingly. The field equations (45) determine that only any one component of the 1-form $^+\omega$ may be a non zero constant. This $^+\omega$ is either time-like or space-like and static and accordingly may be called electrostatic or magnetostatic. Thus, we obtain a series of related complex substratum solutions with different possible constant values of $^-\omega^\pm$ and $^+\omega$. The modulus of the total connection solution $\omega$ is not given by eq. (48). Furthermore, it may be shown that the modulus of any of these related complex solutions increases due to $^+\omega$ and is larger than the value given by this equation.

## Connection Excitations

To obtain field excitations we perform perturbations of the geometric objects in the field equations. Then the linear differential equation for any perturbation of the connection takes the general form

$$\tfrac{2}{\sqrt{-g}}\partial_\rho\left(\sqrt{-g}\,g^{\rho\mu}g^{\alpha\nu}\partial_{[\mu}\delta\omega_{\nu]d}\right) + 4\omega_\rho^c\omega_c^\rho\delta\omega_d^\alpha + {}^1L_{d\rho}^{\alpha c}\delta\omega_c^\rho + {}^2L_d^{\alpha\mu\nu}\delta g_{\mu\nu} = 4\pi\alpha\,\delta J_d^\alpha \ . \tag{51}$$

In particular, we assume that we can obtain a connection excitation solution with algebraic components only in the complex Minkowski plane $K_k$. Then the complex connection and its excitation have the form

$$\omega_\rho^{\hat\alpha} = \left(\omega^\pm\right)_\rho^{\hat\alpha} + \delta\omega_\rho^{\hat\alpha} \ , \tag{52}$$

$$\omega^\pm = -m_g dx^{\hat\alpha\pm}\left(\kappa_{\hat\alpha} \pm i\kappa_{\hat\alpha}\right) \ . \tag{53}$$

The excitation equation becomes, using the complex metric $g$ of the complex Minkowski plane $K_k$ that involves lowering the indices and taking the complex conjugate,

$$^*d\,{}^*d\delta\omega_\delta^\nu + 4\left(\delta\omega_\rho^{\hat\chi}\omega^{\hat\alpha\rho}\omega^{\hat\beta\nu} + \omega_\rho^{\hat\chi}\delta\omega^{\hat\alpha\rho}\omega^{\hat\beta\nu} + \omega_\rho^{\hat\chi}\omega^{\hat\alpha\rho}\delta\omega^{\hat\beta\nu}\right)\times$$
$$\left(g_{\hat\delta\hat\beta}g_{\hat\chi\hat\alpha} - g_{\hat\delta\hat\alpha}g_{\hat\chi\hat\beta}\right) = 4\pi\alpha\,\delta J_\delta^\nu \ . \tag{54}$$

**(References)**

[1] P. W. Higgs, Phys. Lett. **12**, 132 (1964); P. W. Higgs, Phys. Rev. Lett. **13,** 508 (1964).[2]

[2] Gustavo R. González-Martín, Phys. Rev. D **35**, 1225 (1987); Gustavo R. González-Martín, Gen. Rel. and Grav. **22**, 481 (1990).

[3] Hideki Yukawa, Proc. Phys. Math. Soc. (Japan) **17**, 48 (1935).

[4] Gustavo R. González-Martín, Rev. Mexicana de Física **49 Sup. 3**, 118 (2003).

[5] W. Misner, K. Thorne, and J. Wheeler, *Gravitation* (W. H. Freeman and Co., San Francisco,1973), Vol. **1**, p.715.

[6] W. Killing, J. Reine Angew. Math. **109**, 121 (1892).

[7] Gustavo R. González-Martín, arXive physics/0405126

[8] Robert E. Marshak, *Conceptual Foundations of Modern Particle Physics* (World Scientific, Singapore, 1993), p.345.

[9] Gustavo R. González-Martín, arXive physics/0009066; arXive physics/0405097

[10] S. Helgason, *Differential Geometry and Symmetric Spaces* (Academic Press, New York 1962), p.285.

[11] Gustavo R. González-Martín *Physical Geometry* (U. Simón Bolívar, Caracas, 2006).

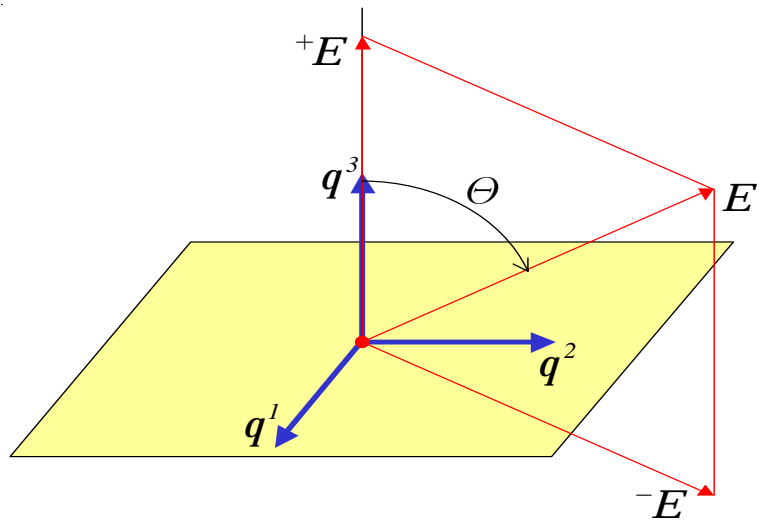

Polar angle $\Theta$ in the su(2)$_Q$ algebra determined by the even and odd generators